\documentclass[
    ,final            % use final for the camera ready runs
  ]
  {aipproc}

\layoutstyle{6x9}

\begin{document}

\title[H$_{2}$ Fluorescence with Partial Frequency Redistribution]{Modeling H$_{2}$ Fluorescence in Planetary Atmospheres with Partial Frequency Redistribution}


\author{R. E. Lupu}{
  address={Dept. of Physics and Astronomy, Johns Hopkins University, Baltiomre, MD 21218}
}

\author{P. D. Feldman}{
  address={Dept. of Physics and Astronomy, Johns Hopkins University, Baltiomre, MD 21218}
}

\author{S. R. McCandliss}{
  address={Dept. of Physics and Astronomy, Johns Hopkins University, Baltiomre, MD 21218}
%  ,altaddress={<author1 address>} % additional visiting address
}

\author{K. France}{
  address={Center for Astrophysics and Space Astronomy, University of Colorado, Boulder, CO 80309}
}

\begin{abstract}
We present the modeling of partial frequency redistribution (PRD) effects for the fluorescent emission lines of molecular hydrogen, the general computational approximations, and the applications to planetary atmospheres, as well as interstellar medium. Our model is applied to $FUSE$ observations of Jupiter, Saturn, and reflection nebulae, allowing an independent confirmation of the H$_{2}$ abundance and the structure of planetary atmospheres. 
\end{abstract}

\maketitle

%%%%%%%%%%%%%%%%%%%%%%%%%%%%%%%%%%%%%%%%%%%%
%% MAINMATTER
%%%%%%%%%%%%%%%%%%%%%%%%%%%%%%%%%%%%%%%%%%%%

\section{PRD Fluorescence Model}

\paragraph{Frequency Redistribution} $FUSE$ observations of planetary atmospheres and reflection nebulae have revealed the need to include PRD in H$_{2}$ fluorescence radiative transfer models. The angle-averaged laboratory frame redistribution function describes the conditional probability that a photon absorbed at x$_{i}$ Doppler widths from the center of line $i$, is emitted at x$_{f}$ Doppler widths from the center of line $f$ \citep{Mihalas:1978}. Complete redistribution (CRD) refers to the photons being re-emitted in the line core according to the Voigt profile, while PRD takes into account the changes in the line profile due to coherent scattering in the line wings. PRD becomes important for fast transitions and integrated optical depths larger than 10$^{3}$. Observable effects include shifts in the peak wavelength in emission due to variations of the exciting spectrum over the absorbing line profile, a decrease in the line-to-continuum contrast, and a decrease in the number of photons scattered in subordinate lines (cross redistribution, XRD).\\[-0.25in]

\paragraph{Radiative Transfer} The PRD radiative transfer problem is more complex than the CRD case, due to the heavy couplings in frequency in addition to spatial correlations and detailed balance \citep{HubenyLites:1995,Uiten:2001}. Even restricting the geometry to the plane-parallel case, and assuming fixed level populations, the problem remains computationally expensive due to intrinsically large optical depths and fine frequency grids. We investigated a set of numerical methods (see Table~\ref{table}) that would allow the full treatment of the H$_{2}$ molecule (10$^{4}$ transitions) to become feasible. We find that the multilayer approximation is the best approach, with a computing time independent of optical depth, and no matrix inversions required. This method is a layer-by-layer extension of the optically thin single layer solution of \citet{LiuDalgarno:1996}. The results are consistently within 10\% of the Feautrier lambda iteration \citep{Mihalas:1978} for grids of at least 10 points per dex in optical depth.
 
\begin{table}[h]
\begin{tabular}{lrrrr}
\hline
  & \tablehead{1}{r}{n}{Lambda Iteration \\(Feautrier solution)}
  & \tablehead{1}{r}{n}{Multilayer}
  & \tablehead{1}{r}{n}{Single\\layer}   \\
\hline
Resources & High & Moderate & Low\\
Consistency & Can be unstable & Constrained & Constrained\\
Convergence & Scales with optical depth & 2-pass layer-by-layer & Single step\\
Spatial variations  & Yes & Yes & No \\
\hline
\end{tabular}
\caption{Characteristics of the radiateve transfer methods used.}
\label{table}
\end{table}

\section{Results}

\begin{figure}[h]
   %\begin{center}
   \begin{tabular}{ccc}
   \includegraphics[angle=90,width=2.in,clip]{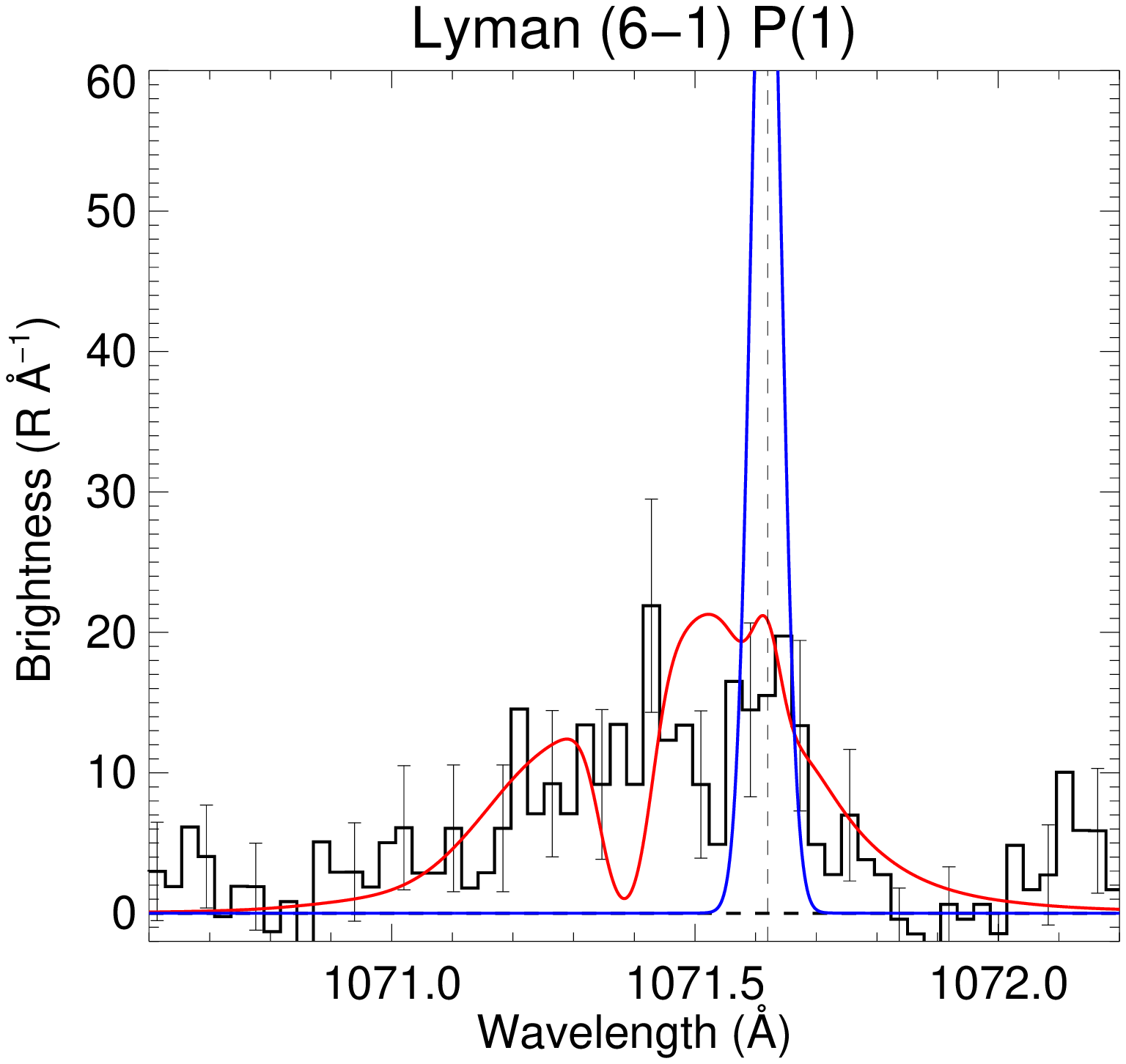}
   \includegraphics[angle=90,width=2.in,clip]{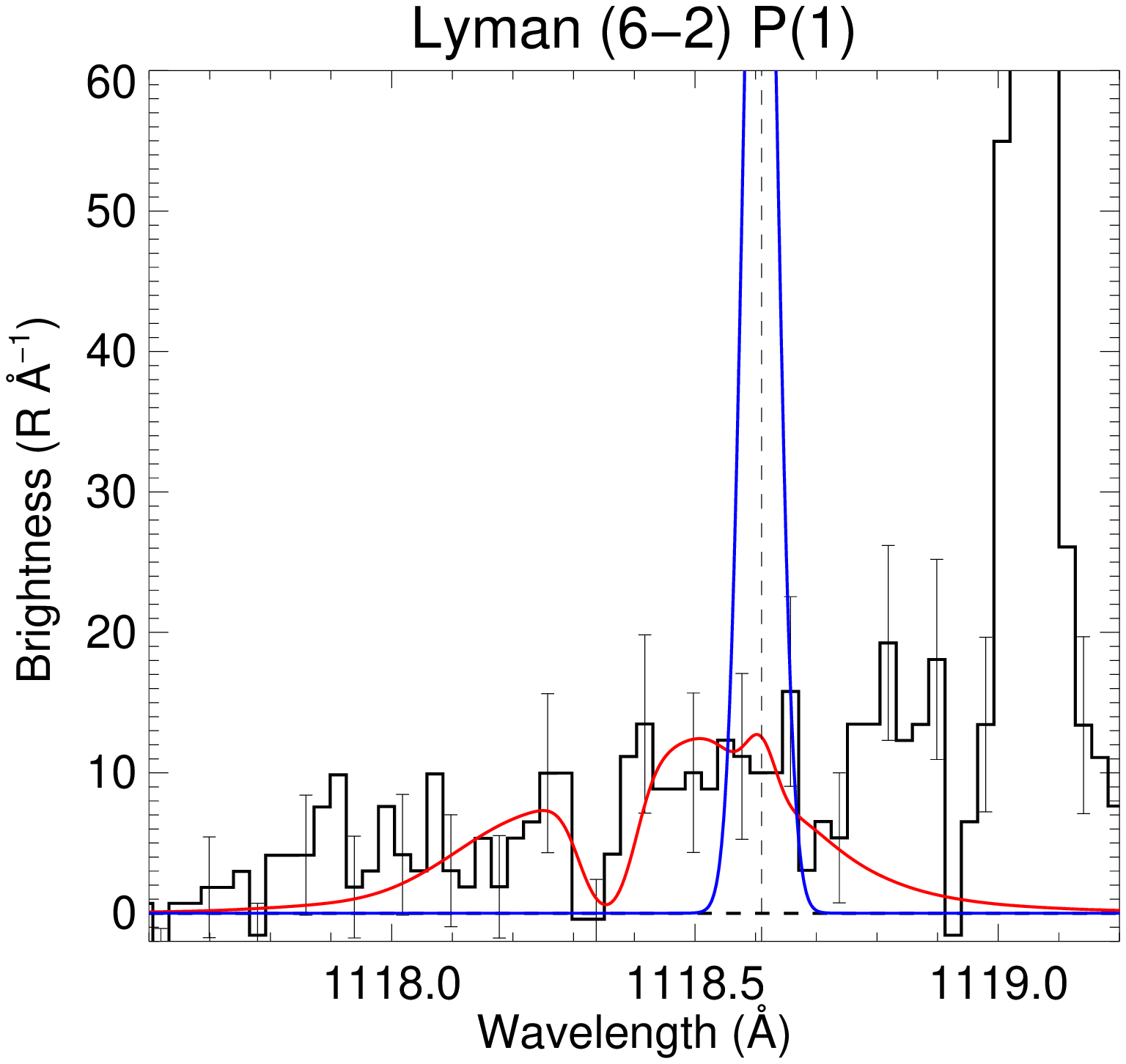}
   \includegraphics[angle=90,width=2.in,clip]{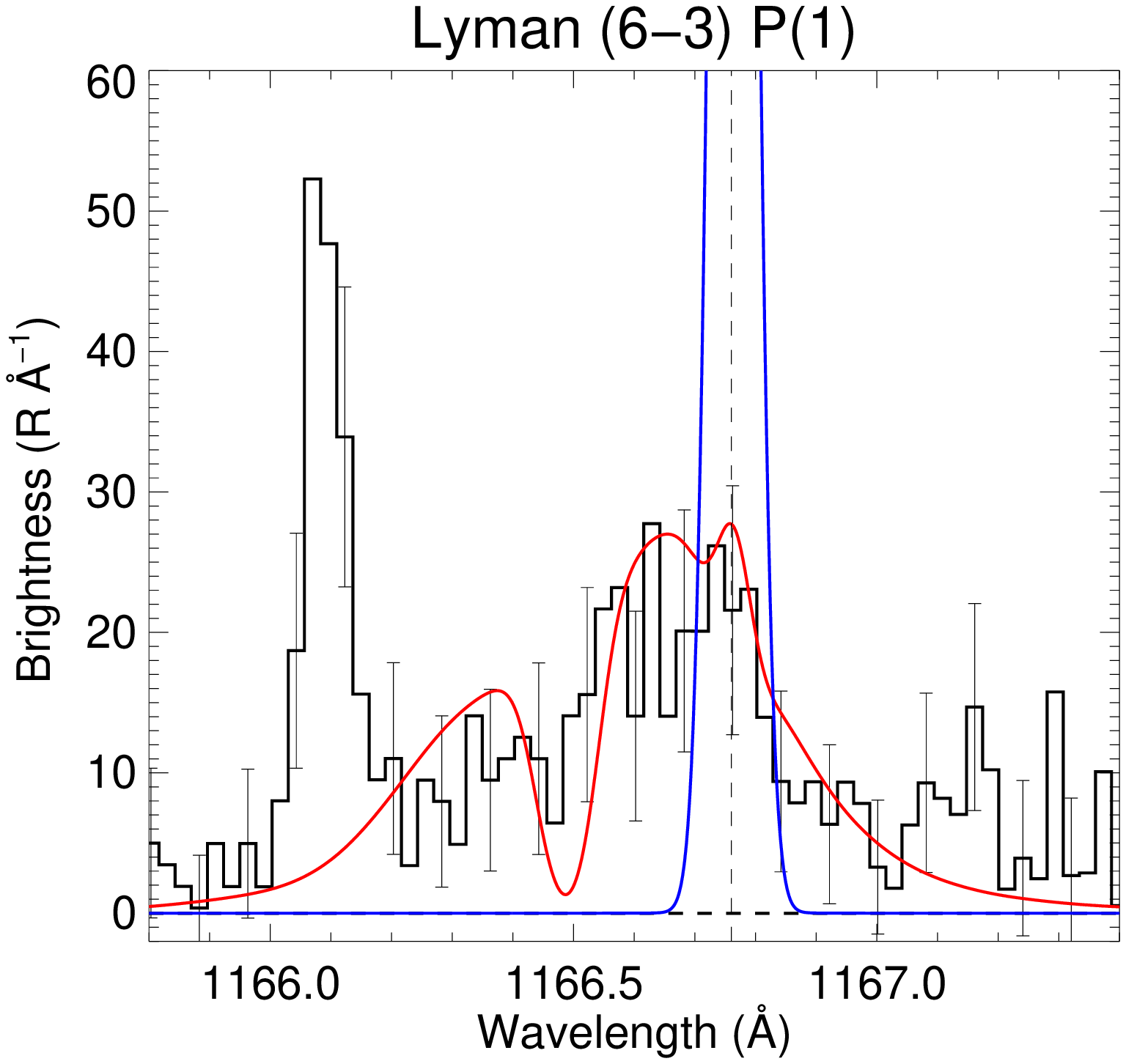}
   \end{tabular}
   %\end{center}
   \caption{ \label{jup} 
$FUSE$ MDRS Jupiter data, FWHM 0.08~\AA\ (black), PRD (red), and CRD (blue) models. }
   \end{figure}

\begin{figure}[h]
   %\begin{center}
   \begin{tabular}{ccc}
   \includegraphics[angle=90,width=2.in,clip]{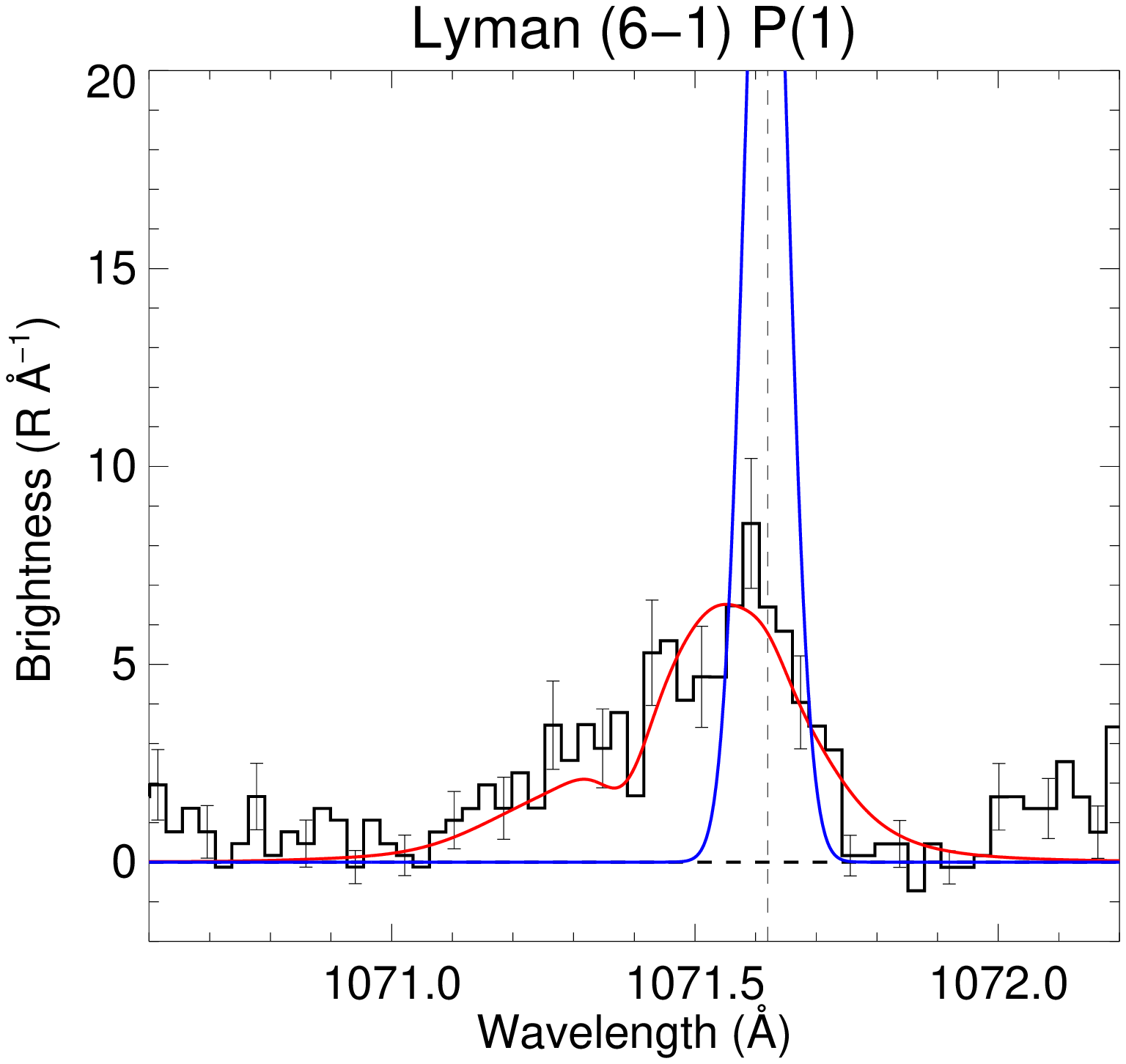}
   \includegraphics[angle=90,width=2.in,clip]{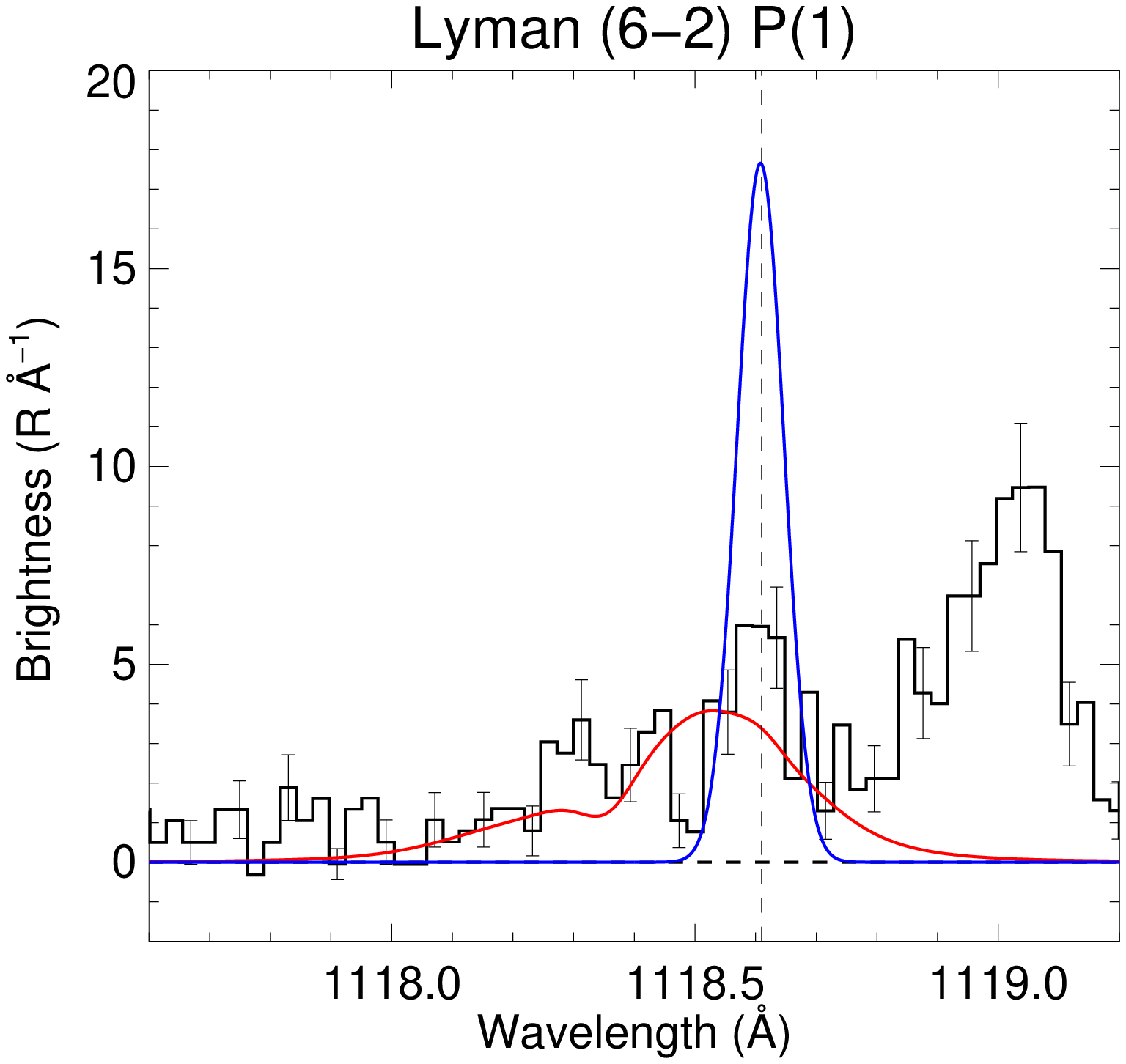}
   \includegraphics[angle=90,width=2.in,clip]{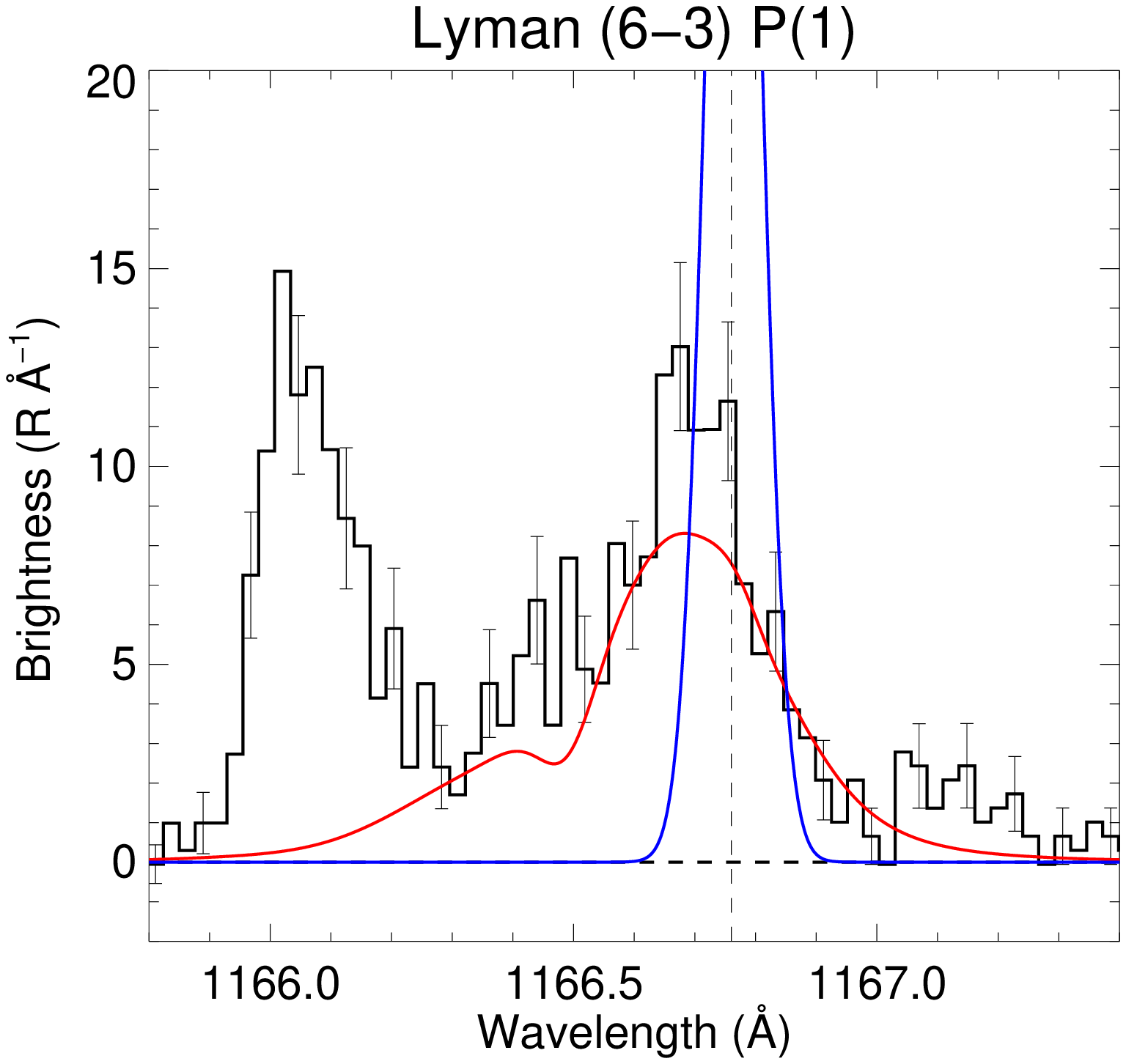}
   \end{tabular}
   %\end{center}
   \caption{ \label{sat} 
$FUSE$ LWRS Saturn data, FWHM 0.12~\AA\ (black), PRD (red), and CRD (blue) models.}
   \end{figure}

\paragraph{Planetary Atmospheres} The PRD effects on resonance lines and overlapping transitions in planetary atmospheres have been discussed previously \citep{Barth:2004,Grif:2000}. The current $FUSE$ observations represent the first account of PRD effects in subordinate molecular fluorescent lines. The lines in the Lyman (6~$-$~v") progression of molecular hydrogen pumped by the solar Ly$\beta$ show a broad asymmetric profile, consistent with the shape of the overlap between the (6~$-$~0) P(1) line and Ly$\beta$, and emphasize the importance of coherent scattering in the line wings. The $FUSE$ spectra of the Jupiter limb and Saturn disk shown in Figure~\ref{jup} and Figure~\ref{sat}, respectively, are well-fitted by the XRD model (red). For comparison, the CRD model is also shown in blue. The synthetic spectra have been obtained by integrating the full atmospheric models, and using the multilayer approximation.

 % \hspace{-2cm}

\begin{figure}[h]
  \includegraphics[height=1.8in]{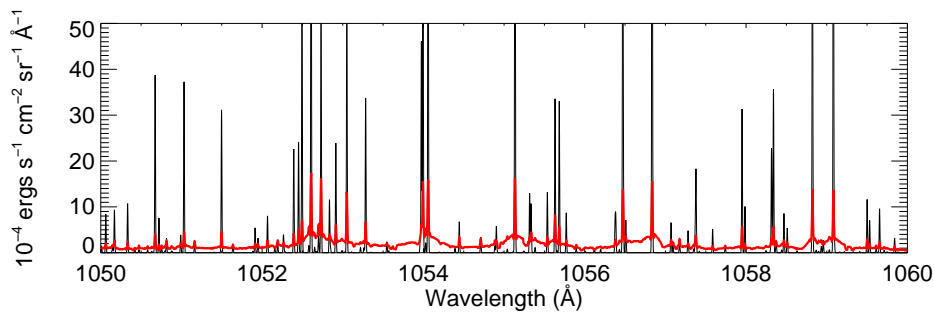}
  \caption{Comparison between the H$_{2}$ spectrum predicted by the CRD (black) and PRD (red) models. }
  \label{2023}
\end{figure}

\paragraph{NGC 2023}In spite of predictions based on the infrared H$_{2}$ observations \citep{Takami:2000}, high excitation non-thermal H$_{2}$ absorption detected in the UV \citep{Meyer:2001}, and UV H$_{2}$ emission at 1575$-$1608\AA\ observed by HUT, the molecular hydrogen lines in the $FUSE$ bandpass do not match the CRD model predictions. Here we test the assumption that this discrepancy is due to the effects of PRD and XRD. We employ a toy model, with an incident 22,000~K blackbody spectrum normalized to 5000 times Harbig galactic mean \citep{Meyer:2001}, illuminating a uniform H$_{2}$ cloud with rotational temperature of 1500~K, Doppler $b$ parameter of 1.8~km~s$^{-1}$, and an integrated column of 5$\times$10$^{20}$~cm$^{-2}$. The radiative transfer was performed in a single layer approximation, to minimize computational resources. The results shown in Figure~\ref{2023} represent the first XRD calculation of a fluorescent molecular spectrum. This exercise shows that the PRD effects will decrease the line contrast, making the H$_{2}$ features less prominent at high spectral resolution. In future work, we expect stronger peak suppression in a multilayer treatment due to the effects of multiple scattering.

%%%%%%%%%%%%%%%%%%%%%%%%%%%%%%%%%%%%%%%%%%%%%%%%
%% BACKMATTER
%%%%%%%%%%%%%%%%%%%%%%%%%%%%%%%%%%%%%%%%%%%%%%%%

\begin{theacknowledgments}

We thank Prof. D. Strobel for providing the atmospheric models of Jupiter and Saturn, and the FUSE ground system personnel for planning and executing the observations. The data was obtained for the Guaranteed Time Team by the NASA-CNES-CSA $FUSE$ mission operated by the Johns Hopkins University. Financial support was provided by NASA contract NAS5-32985, and NASA grants NAS5-13085 and NAG5-13719. 
\end{theacknowledgments}



\bibliographystyle{aipproc}   % if natbib is available

\bibliography{fusebib}

\IfFileExists{\jobname.bbl}{}
 {\typeout{}
  \typeout{******************************************}
  \typeout{** Please run "bibtex \jobname" to optain}
  \typeout{** the bibliography and then re-run LaTeX}
  \typeout{** twice to fix the references!}
  \typeout{******************************************}
  \typeout{}
 }

\end{document}